\documentclass[twocolumn]{aastex62}
\usepackage{amsmath}
\usepackage{color}
\usepackage{amssymb}
\usepackage{float}
\addcontentsline{toc}{section}{Acknowledgement}

\usepackage{graphicx}

\def\apjs{ApJS}
\def\apj{ApJ}
\def\apjl{ApJ Lett}
\def\aj{AJ}
\def\aap{A\&A}

\def\mnras{MNRAS}
\def\nat{Nature}

\def\Mu5{\dot{\cal{M}}}

\begin{document}

\title{Cosmological constraints from the redshift dependence of the
Alcock-Paczynski effect: Possibility of estimateing the non-linear systematics using fast simulations}


\author{Qinglin Ma}
\affiliation{School of Physics and Astronomy, Sun Yat-Sen University, Guangzhou 510297, P.R.China}
\author{Yiqing Guo}
\affiliation{School of Physics and Astronomy, Sun Yat-Sen University, Guangzhou 510297, P.R.China}
\author{Xiao-Dong Li}
\affiliation{School of Physics and Astronomy, Sun Yat-Sen University, Guangzhou 510297, P.R.China}
\author{Xin Wang}
\affiliation{School of Physics and Astronomy, Sun Yat-Sen University, Guangzhou 510297, P.R.China}
\author{Haitao Miao}
\affiliation{School of Physics and Astronomy, Sun Yat-Sen University, Guangzhou 510297, P.R.China}
\author{Zhigang Li}
\affiliation{Nanyang Normal University, 1638 Wolong Rd, Wolong, Nanyang, China}
\author{Cristiano G. Sabiu}
\affiliation{Department of Astronomy, Yonsei University, Seoul, Korea}
\author{Hyunbae Park}
\affiliation{Kavli Institute for the Physics and Mathematics of the Universe (WPI), The University of Tokyo Institutes for Advanced Study, The University 
of Tokyo, Kashiwa, Chiba 277-8583, Japan}

\correspondingauthor{Xiao-Dong Li, Haitao Miao}

\begin{abstract}
The tomographic AP method is so far the best method in separating the Alcock-Paczynski (AP) 
signal from the redshift space distortion (RSD) effects and deriving powerful constraints 
on cosmological parameters using the $\lesssim40h^{-1}\ \rm Mpc$ clustering region.
To guarantee that the method can be easily applied to the future large scale structure (LSS) surveys, 
we study the possibility of estimating the systematics of the method using fast simulation method.
The major contribution of the systematics comes from the non-zero redshift evolution of the RSD effects, 
which is quantified by $\hat\xi_{\Delta s}(\mu,z)$ in our analysis,
and estimated using the BigMultidark exact N-body  simulation and approximate COLA  simulation samples. 
We find about 5\%/10\% evolution when comparing the $\hat\xi_{\Delta s}(\mu,z)$  measured as $z=0.5$/$z=1$ to the measurements at $z=0$.
We checked the inaccuracy in the 2pCFs computed using COLA,
and find it 5-10 times smaller than the intrinsic systematics of the tomographic AP method, 
indicating that using COLA to estimate the systematics is good enough.
Finally, we test the effect of halo bias, and find
$\lesssim$1.5\%  change in $\hat\xi_{\Delta s}$ when varying the halo mass within the range of
$2\times 10^{12}$ to $10^{14}$ $M_{\odot}$.
We will perform more studies to achieve an accurate and efficient estimation of 
the systematics in redshift range of $z=0-1.5$. 
\end{abstract}


\keywords{}.

\section{Introduction}\label{sec:intro}

The large-scale structure (LSS) of the Universe encodes an
enormous amount of information about its expansion and structure growth histories.
In the past two decades, large-scale surveys of galaxies
have greatly enriched our understanding about the Universe
\cite{2df:Colless:2003wz,beutler20116df,blake2011wigglez,blake2011wigglezb,york2000sloan,Eisenstein:2005su,Percival:2007yw,
anderson2012clustering,alam2017clustering}.
The next generation of LSS surveys, such as 
DESI\footnote{https://desi.lbl.gov/}, EUCLID\footnote{http://sci.esa.int/euclid/}, 
LSST\footnote{https://www.lsst.org/},
WFIRST\footnote{https://wfirst.gsfc.nasa.gov/}
will enable us to measure 
the $z\lesssim1.5$ Universe to an unprecedented precision,
shedding light on the dark energy problem \citep{riess1998observational,perlmutter1999measurements,
weinberg1989cosmological,miao2011dark,weinberg2013observational}.



The Alcock-Paczynski (AP) test \citep{Alcock} is a pure geometric probe 
of the cosmic expansion history based on the comparison of observed
radial and tangential sizes of objects that are known to be isotropic. 
Under a certain cosmological model, the radial and tangential sizes of some distant objects or structures take the forms of 
$\Delta r_{\parallel} = \frac{c}{H(z)}\Delta z$ and $\Delta r_{\bot}=(1+z)D_A(z)\Delta \theta$, where $\Delta z$, $\Delta \theta$ are their observed redshift span and angular size, while $D_A$, $H$ are the angular diameter distance and the Hubble parameter computed using theories.
In the case that incorrect models were assumed for computing $D_A$ and $H$,
the values of $\Delta r_{\parallel}$ and $\Delta r_{\bot}$ are wrongly estimated, 
resulting in geometric distortions along the line-of-sight (LOS) and tangential directions. 
This distortion can be quantified via statistics of the large-scale galaxy distribution, and has been widely used in galaxy surveys to place constraints on the cosmological parameters
\citep{ryden1995measuring,Ballinger1996,matsubara1996cosmological,outram20042df,blake2011wigglez,lavaux2012precision,alam2017clustering,
Qingqing2016,KR2018}.






The tomographic AP\citep{LI14,LI15} method is a novel application of the original AP test 
that uses the {\it redshift evolution} of the clustering anisotropies, which are sensitive to the AP effect while being relatively insensitive to the anisotropies 
produced by redshift space distortions (RSD).
This makes it possible to differentiate the AP distortion from the large contamination from the RSD effect.
\cite{LI16} firstly applied the method to the 
SDSS (Sloan Digital Sky Survey) BOSS (Baryon Oscillation Spectroscopic Survey) DR12 galaxies,
and achieved $\sim30-40\%$ improvements in the constraints 
on the ratio of dark matter $\Omega_m$ and dark energy equation of state (EOS) $w$
when combining the method with the datasets of 
the Planck measurements of Cosmic Microwave Background (CMB) \citep{planck2016ade}, 
type Ia supernovae (SNIa) \citep{betoule2014},
baryon acoustic oscillations (BAO) \citep{Anderson2014}
and local $H_0$ measurement\citep{Riess2011,Efstathiou2014}. 
In follow-up studies, \cite{LI18,Zhang2019} studied 
the constraints on models with possible time-evolution of dark energy EOS,
and showed that the method can decrease the errors of the parameters by $40-50$\%.
\cite{LI19} forecast the performance of the tomographic AP method on future DESI data and found that
when combining with Planck+BAO datasets the dynamical dark energy constraints are improved by a factor of 10. 

The tomographic AP method is so far the best method in separating the AP 
signal from the RSD effects and extract information from the 
$\lesssim 40h^{-1}\ \rm Mpc$ clustering regions.
Thus, it is essentially important to conduct more studies
to guarantee that the method can be applied to future LSS surveys.
The biggest caveat of the method is the systematics from the time evolution of the RSDs.
On the scales explored by the tomographic AP method, the accurate quantification of RSDs can only be achieved via
precise numerical simulations.

\cite{LI16} utilized the Horizon Run 4 (HR4) N-body simulation \citep{kim2015horizon} to 
estimate the systematics when applying the method to the SDSS galaxies.
They showed that, within the redshift range of 0.2-0.5/0.2-0.7, 
the systematics creates $\lesssim2\%/6\%$ time evolution in the anisotropic correlation function $\hat \xi_{\Delta s}(\mu)$.
\cite{LI18} conducted more tests on the systematics and 
found that, the shifts in the derived cosmological parameters caused by the systematics 
are well below the $1\sigma$ statistical error.
\cite{Park:2019mvn} studied the systematics 
using simulations generated in five different cosmologies,
and reported a non-negligible cosmology  dependence.
More studies about the systematics have been performed in \cite{LI15,LI19},
and their conclusions are consistent with \cite{LI16,LI18,Park:2019mvn}.

While many studies on the systematics of the method have been performed, 
it is necessary to conduct more checks regarding the systematics including:
\begin{itemize}
 \item Firstly, the redshift coverage of current studies of the method are limited to $z\lesssim0.7$.
To meet the requirements of future surveys, this should be enlarged to  $z\sim1.5$.
 \item Secondly, so far the systematics are mainly studied by using  the Horizon Run N-body simulations \citep{Kim2011,kim2015horizon},
 which are performed in the $\Omega_m =0.26$ $\Lambda$CDM cosmology.
 In different cosmologies, the influence of the RSD effect is also different.
 The cosmological dependence of the systematics is firstly studied in \cite{Park:2019mvn},
 and remains to be investigated in more details.
 \item Finally, it would be necessary to have a fast method for the systematics estimation. 
 The next generation experiment will survey an unprecedented large volume of the Universe.
 A fast and accurate method would be very helpful if we were to 
 estimate the systematics of the tomographic AP method for these surveys.
\end{itemize}

In this paper, we explore the first and third issues listed above. 
In Section 2, we introduce the simulation materials used in the paper,
including both the N-body and fast simulation samples. 
The methodology of the tomographic AP method is briefly reviewed in Section 3.
In Section 4 we present our results, including the high-$z$ measurements of the systematics,
a comparison between N-body and fast simulation results,
a test on the clustering scales,
and a check on the halo bias effect.
We conclude in Section 5.

\section{Simulation} \label{sec:simu}

We use two sets of simulation samples, the BigMultiDark (BigMD) N-body simulation sample \citep{BD}, 
and also a set of fast simulation samples generated using the COLA (COmoving Lagrangian Acceleration) algorithm. 

In both sets of samples, to mimic the redshift-space distortions (RSD) caused by galaxy peculiar velocities,
we perturb the positions of halos along the radial direction,
using the following formula
\begin{equation}\label{eq:zvpeu}
\Delta z = (1+z) \frac{v_{{\rm LOS}}}{c},
\end{equation}
where $v_{\rm LOS}$ is the line-of-sight (LOS) component of the peculiar velocity of galaxies, and $z$ is the cosmological redshift.

For testing the redshift-evolution of the RSD effect,
in both samples we use the outputs of halos at 12 redshifts, 
i.e. $z=$\{0, 0.102, 0.2013, 0.2947, 0.4037, 0.5053, 0.6069, 0.7053, 0.7976, 0.8868, 1, 1.445\},
respectively.



\subsection{The BigMD simulation}
  
The Multiverse simulations are a set of cosmological N-body simulations designed to 
study the effects of cosmological parameters on the clustering and evolution of cosmic structures. 
Among them, the BigMultidark simulation is produced using $3\,840^3$ particles in a volume of $(2.5h^{-1}\rm Gpc)^3$
assuming a $\Lambda$CDM cosmology with
 $\Omega_m = 0.307115$, $\Omega_b = 0.048206$, $\sigma_8 = 0.8288$, $n_s = 0.9611$, and $H_0 = 67.77\ {\rm km}\ s^{-1} {\rm Mpc}^{-1}$
\citep{BD}.
The size and number of particles of this simulation is capable for 
the purpose of study in this work (although it is smaller than 
the Horizon Run N-body simulations \citep{kim2009horizon,kim2015horizon} used in \cite{LI14,LI15,LI16}).
Having both a large volume and a good resolutions,
this simulation is able to accurately reproduce the observational statistics of the current redshift surveys \citep{Torres}.

%
\begin{table*}
	\caption{Cosmology parameters of BigMD ang COLA simulation}	
	\begin{center}
		 \setlength{\tabcolsep}{6mm}{	 	
		\begin{tabular}{cccccccc}
			\hline
			Simulation & Box & particles &  $h$ & $\Omega_{m}$ & $\Omega_{b}$ & $n_s$ & $\sigma_{8}$\\
			\hline
			BigMultiDark & 2.5 & $3840^{3}$ & 0.6777 & 0.307115 & 0.048206 & 0.96 & 0.8288 \\
			\hline
			COLA & 0.6 & $600^3$ & 0.6777 & 0.307115 & 0.048206 & 0.96 & 0.8288\\
			\hline
		\end{tabular}}
		\label{bigmd data}
	\end{center}
\end{table*}

We use the public available halo catalogue of BigMD simulation, created using the \textsc{ROCKSTAR}  halo finder \citep{ROCKSTAR}.
\textsc{ROCKSTAR} is a halo finder based on adaptive  hierarchical refinement  of  friends-of-friends  groups  
in six phase-space dimensions  and  one time dimension, 
allowing for robust tracking of substructure.
To make the samples at different redshifts comparable to each other,
we maintain a constant number density $\bar n=$0.001 $(h^{-1}\rm Mpc)^{-3}$ in all snapshots. 
Both halos and subhalos are included in the analysis.
In Appendix A, a rough comparison between the 2pCFs of the BigMD \textsc{ROCKSTAR} halos and the BOSS CMASS galaxies
shows that the simulation sample can recover the observational results at a $\lesssim5\%$ accuracy level.

\subsection{The COLA samples}

Although being powerful in constructing high-quality, realistic dark matter halo catalogues,
N-body simulations are computationally expensive \citep{Angulo12, Fosalba15, Heitmann15, Potter17},
making it difficult to be used for creating a large number of mock catalogs for current and future large surveys such as SDSS, LSST, Euclid and DESI.
In order to circumvent this problem, some fast algorithms have been proposed to
reproduce the large-scale statistics of N-body simulations.
An incomplete list of these algorithms includes PTHalos \citep{Manera13, Manera15}, PATCHY (PerturbAtion Theory Catalog generator of Halo
and galaxY distribution) \citep{Kitaura14}, QPM (Quick Particle Mesh) \citep{White14},
EZmock \citep{Chuang15}, HALOgen \citep{Avila2014} 
and COLA (COmoving Lagrangian Acceleration) \citep{Tassev13}.
These algorithms enable us to study the clustering properties of the LSS in an efficient manner \citep{Chuang14}.


In this work, we concentrate on the possibility of using the COLA algorithm \citep{Tassev13}
as a replacement for N-body method to quickly generate large number of mocks
for the estimation of systematics in the tomographic AP method.
The second order Lagrangian perturbation Theory (2LPT) adopted by the COLA code is fast in computation and still 
accurate enough in describing the large scale dynamics.
Due to the fact that  2LPT can be easily incorporated in any N-body code,
COLA combines it within N-body simulations by adopting the
2LPT for time integration for large scale dynamical evolution, 
and using a full-blown N-body code with Particle-Mesh (PM) algorithm to deal with small scale dynamics. 
Compared with the fastest simulation algorithms, COLA is better in simulating the structures on non-linear scales \citep{Chuang14},
which is rather suitable for the science case in this work.



We generate 150 COLA simulations in the BigMD cosmology as reference samples to the BigMD simulation. 
For each COLA simulation, $600^3$ particles in a boxsize of 512 $h^{-1}\rm Mpc$ are used and the timesteps are set at 20.
The resolution of the COLA simulations are 30\% higher than the BigMD simulation,
and the total volume of the 150 simulations, being $\approx(2\,720 h^{-1}\rm Mpc)^3$,
is also slightly larger than the BigMD boxsize.
Besides, to be comparable with the BigMD simulation samples, 
we still use \textsc{ROCKSTAR} halo finder 
to build up the halo catalogues from the COLA particles.

%



\section{Methodology} \label{sec:method}

In what follows, we first briefly introduce the AP effect and its redshift evolution,
and then discuss how to use it for estimating cosmological parameter, 
as well as the necessity of conducting studies about its systematics.

\subsection{The AP Effect} 

The AP effect \citep{Alcock} is known as geometric distortions 
when incorrect cosmological models are assumed for transforming redshift to comoving distance
(see \cite{LI14} for a detailed description). 
We probe the size of an object in the universe
through measuring the redshift span $\Delta z$ and angular size $\Delta \theta$,  
which are represented as,
\begin{equation}
	\begin{aligned}
	&\Delta r_{\parallel} = \frac{c}{H(z)}\Delta z, \\
	&\Delta r_{\perp} = (1+z)D_A(z)\Delta \theta,
	\end{aligned}
	\label{1}
\end{equation}
where the cosmological dependence enters via the Hubble parameter $H$ and the angular diameter distance $D_A$. 
In the special case of a flat universe with a constant dark energy EOS, they take the forms of,
\begin{equation}
	\begin{aligned}
	&H(z)=H_{0} \sqrt{\Omega_{m} a^{-3}+\left(1-\Omega_{m}\right) a^{-3(1+w)}},\\
	&D_{A}(z)=\frac{c}{1+z} r(z)=\frac{c}{1+z} \int_{0}^{z} \frac{d z^{\prime}}{H\left(z^{\prime}\right)},
	\end{aligned}
	\label{2}
\end{equation}
where we have neglected the contribution from the radiation. 
Here $a=\frac{1}{z+1}$ is the cosmic scale factor,
$H_{0}$ is the present value of Hubble parameter, and $r(z)$ is the comoving distance.

When we adopted wrong cosmological parameters, 
the $\Delta r_{\parallel}$ and $\Delta r_{\perp}$ in equation \ref{1} would take wrong values, 
resulting in the distorted shape (AP effect) and the wrongly estimated volume (volume effect). 
In the directions parallel and perpendicular to the LOS,
the distortions are 
\begin{equation}
	\begin{aligned}
	&\alpha_{\parallel}(z) = \frac{H_{\rm true}(z)}{H_{\rm wrong}(z)}, \\
	&\alpha_{\perp}(z) = \frac{D_{A,\rm wrong}(z)}{D_{A,\rm true}(z)},
	\end{aligned}
\end{equation}
where ``true'' and ``wrong'' denote the values of quantities in the true and incorrectly assumed cosmologies, respectively.
As a result, apparently the shape of the object is changed by a ratio of,
\begin{equation}
\frac{[\Delta r_{\parallel} /\Delta r_{\perp}]_{\rm wrong}}{[\Delta r_{\parallel} /  \Delta r_{\perp}]_{\rm true}}=\frac{[D_A(z)H(z)]_{\rm true}}{[D_A(z)H(z)]_{\rm wrong}},
\end{equation}
while its volume being changed by
\begin{eqnarray}
\frac{\rm Volume_{\rm wrong}}{\rm Volume_{\rm true}}
&=&
\frac{[\Delta r_{\parallel}(\Delta r_{\perp})^{2}]_{\rm wrong}}
{[\Delta r_{\parallel}(\Delta r_{\perp})^{2}]_{\rm true}}\nonumber\\
&=&
\frac{[D_A(z)^{2}/H(z)]_{\rm wrong}}{[D_A(z)^{2}/H(z)]_{\rm true}},
\end{eqnarray}
respectively.

The above relationships mean that the AP and volume effect, once detected {\it at any clustering scales},
would lead to constraints on the cosmological parameters that control the cosmic expansion.

\begin{figure*}
	\centering
	\includegraphics[width=1\textwidth]{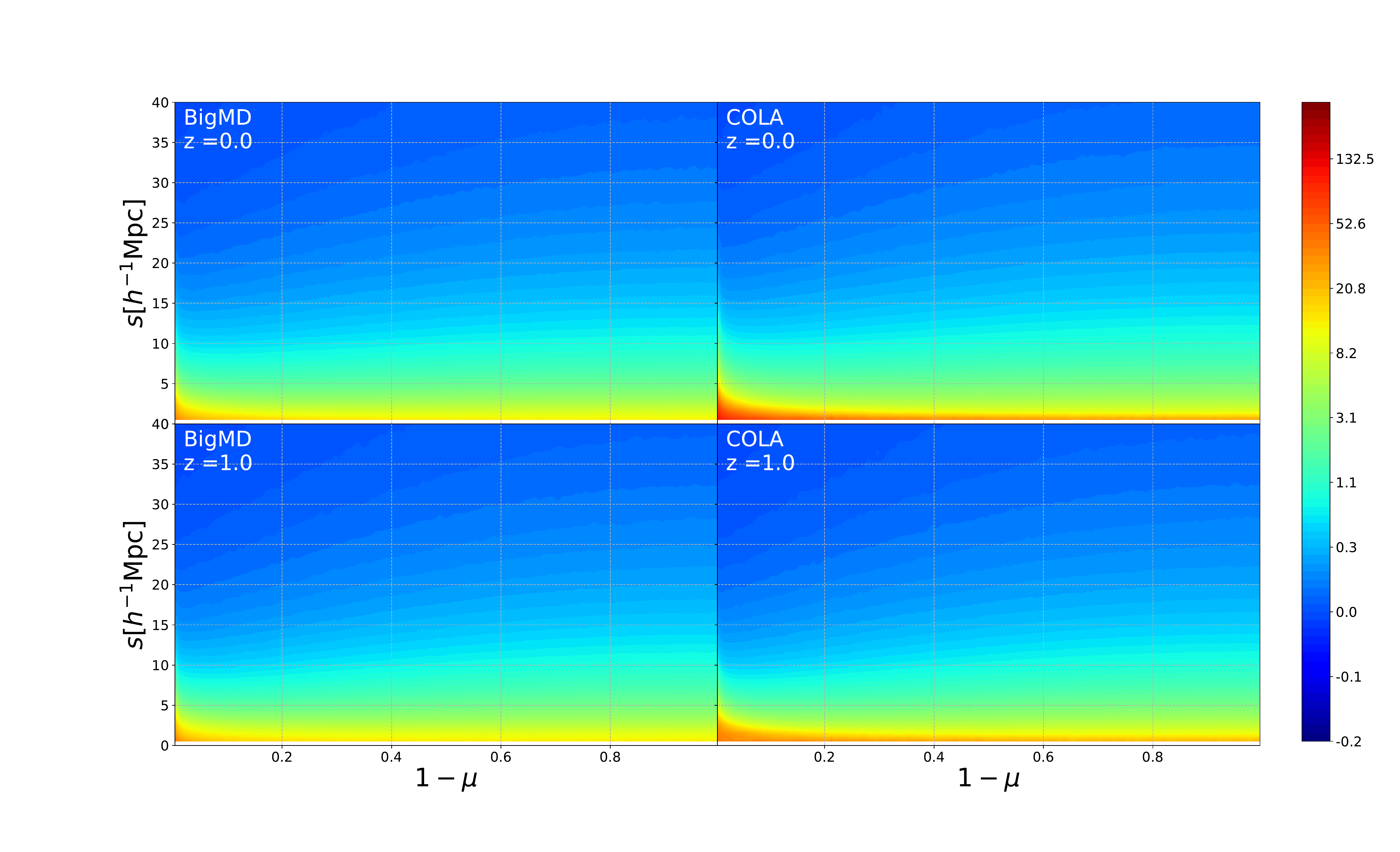}
	\caption{The 2D counter of 2pCF $\xi(s,\mu)$ from the BigMD and COLA samples, 
	at redshift 0 and 1. 
	The contour lines are not horizontal due to the anisotropy produced by the RSD effect. 
	The tilts on the left side and the right side are caused by the 
	FOG (finger-of-god) and Kaiser effects, respectively. 
	The slight difference, which are below a few percent between the BigMD and COLA results, 
	indicates that the COLA has the ability to reproduce the 
	clustering pattern at $\gtrsim5 h^{-1}\rm Mpc$. 
	Noted that the similarity in the region of 
	$1-\mu\lesssim0.2$ implies that
	COLA can correctly simulate the FOG effect.}
	\label{contour}
\end{figure*}

\begin{figure*}
  \centering
  \includegraphics[width=1\textwidth]{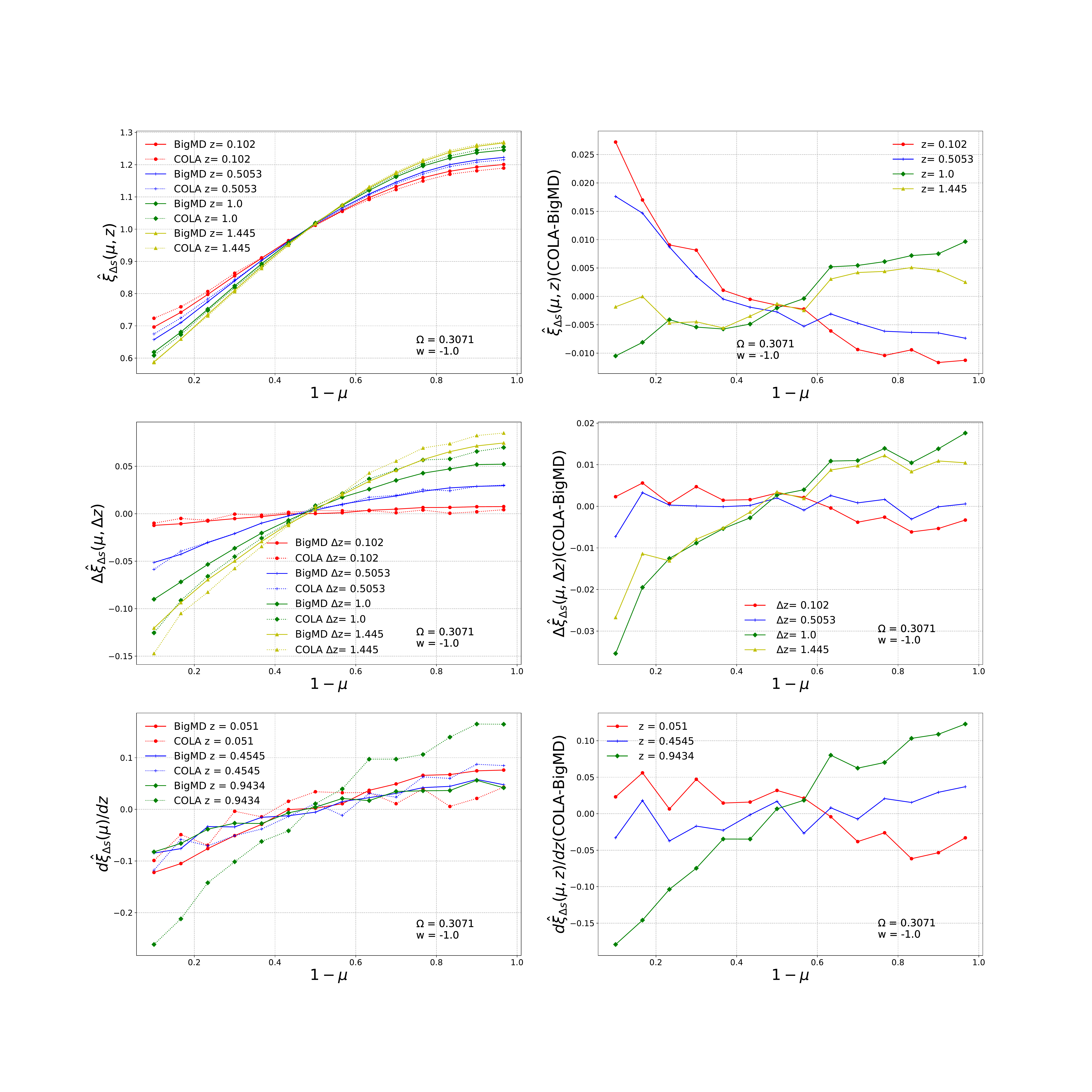}
  \caption{These compare $\hat\xi_{\Delta s}(\mu)$ 
  measured from COLA and BigMD samples when adopting the underlying
  true cosmology $\Omega_m=0.3071$, $w=-1.0$, obtained by integrating
  $\xi(s,\mu)$ within the range $6h^{-1}{\rm Mpc}\le s \le 40h^{-1}\rm Mpc$ 
  together with an amplitude normalization. 
  We split the angular range of $0.06 \le 1 - \mu \le 1$ 
  into as many as 40 bins.
  The upper-left panel shows $\hat \xi_{\Delta s}$ measured at 
  $z=0,\ 0.102,\ 0.5053,\ 1.0,\ 1.445$. 
  Due to the RSD effect as well as its redshift evolution, 
  all curves have $\approx$30\% tilt and are more tilted at higher redshift. 
  In the upper-right panel we find the results from COLA and BigMD 
  merely have a $\lesssim 1\%$ difference at $z<1$, 
  and  $\lesssim 1.5\%$ at $z>1$.
  In the middle panels we plot the evolution against 
  $z=0$ to better quantify the redshift evolution of 
  the clustering anisotropy. 
  In the region of small/large $1-\mu$, 
  the evolution becomes as large as 5\% at $z=0.5$, 
  and increases to 10\%  at $z=1,\ 1.445$. 
  COLA can well reproduce the BigMD results at levels of 
  $\lesssim0.5\%$ except the case of $z=1$, 
  where we find a 2-3\% discrepancy. 
  In the lower panels, we plot ${d \hat \xi_{\Delta s}(\mu)} /{ dz}$ 
  at the three redshifts of $z=0.051,\  0.4545,\ 0.9434$ 
  to measure the redshift evolution from a different perspective. 
  The estimation from COLA has $\lesssim5\%$ error at $z<0.5$, 
  and has a relatively large error of $15\%$ at $z\sim1$.}
  \label{2pcf}
\end{figure*}

\begin{figure*}
  \centering
  \includegraphics[width=0.85\textwidth]{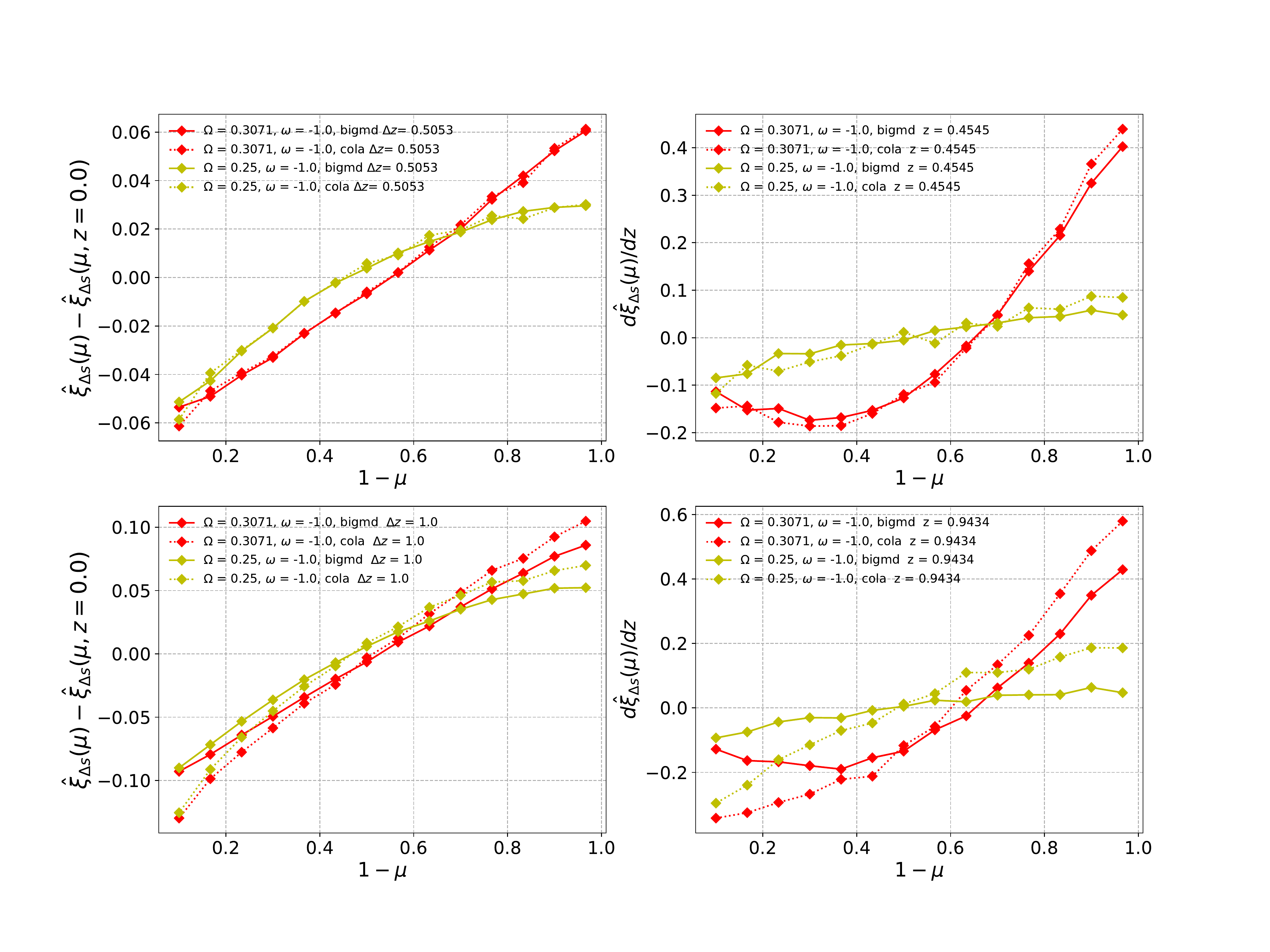}
  \caption{$\Delta \hat \xi_{\Delta s}$ 
  	and $d \hat \xi_{\Delta s} / dz$ measured from the BigMD and COLA samples,
  	in the ``true'' cosmology as well as a wrong cosmology $\Omega_m=0.25$, $w=-1.0$.
  	In the wrong cosmology, the amount of difference between COLA and BigMD results 
  	is similar to what measured in the correct cosmology,
  	while this amount of difference is smaller than the shift of $\hat \xi_{\Delta s}$ and $d \hat \xi_{\Delta s} / dz$ induced 
  	by the wrong cosmology.}
  \label{wrongcos}
\end{figure*}

\begin{figure*}
  \centering
  \includegraphics[width=1\textwidth]{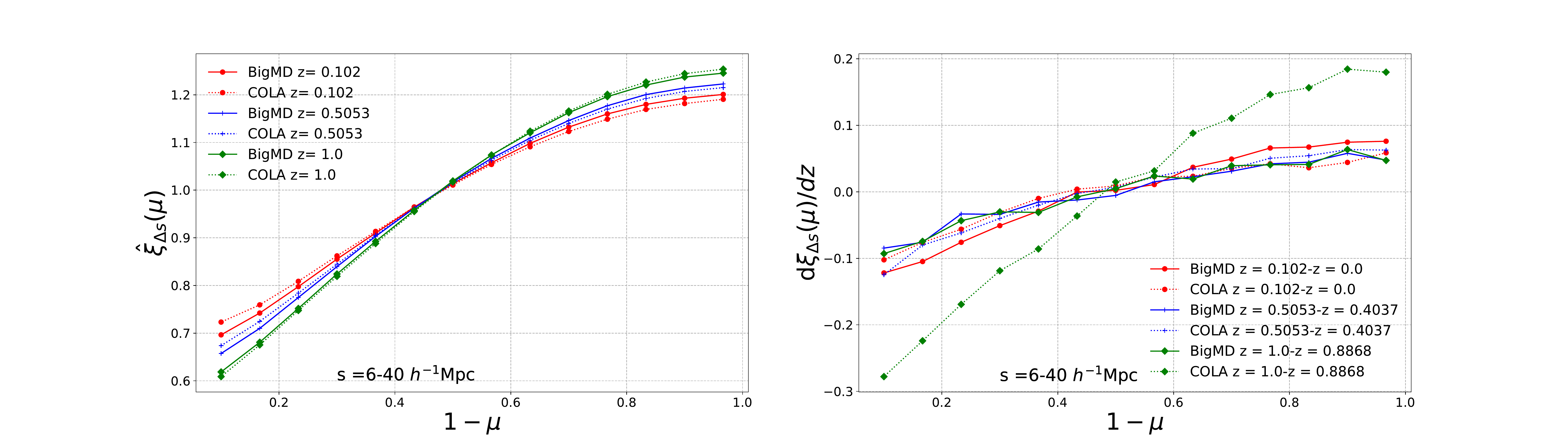}
  \includegraphics[width=1\textwidth]{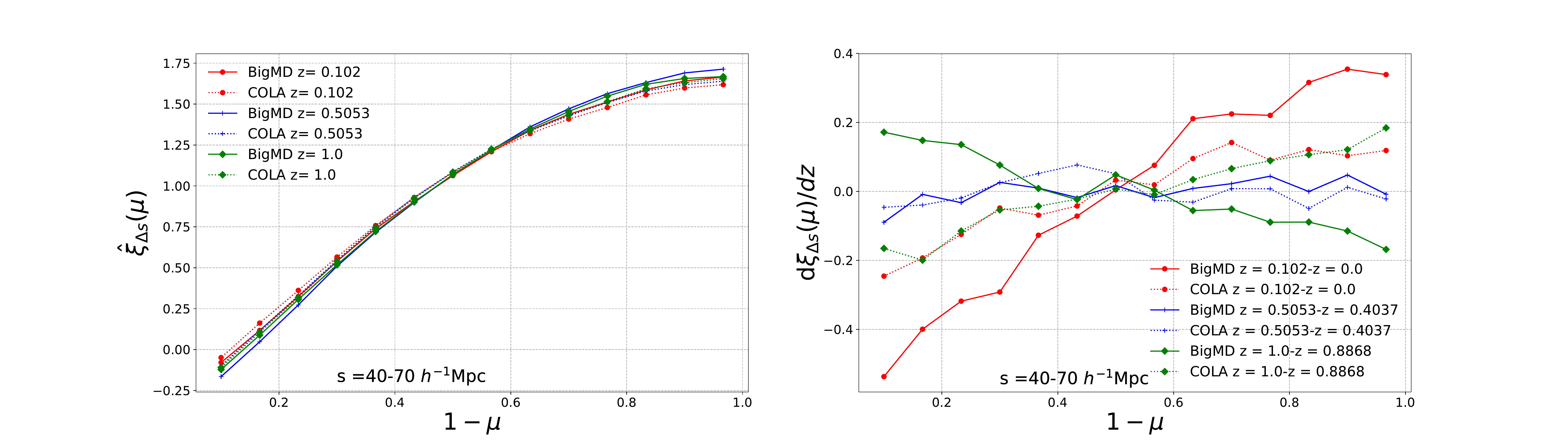}
  \includegraphics[width=1\textwidth]{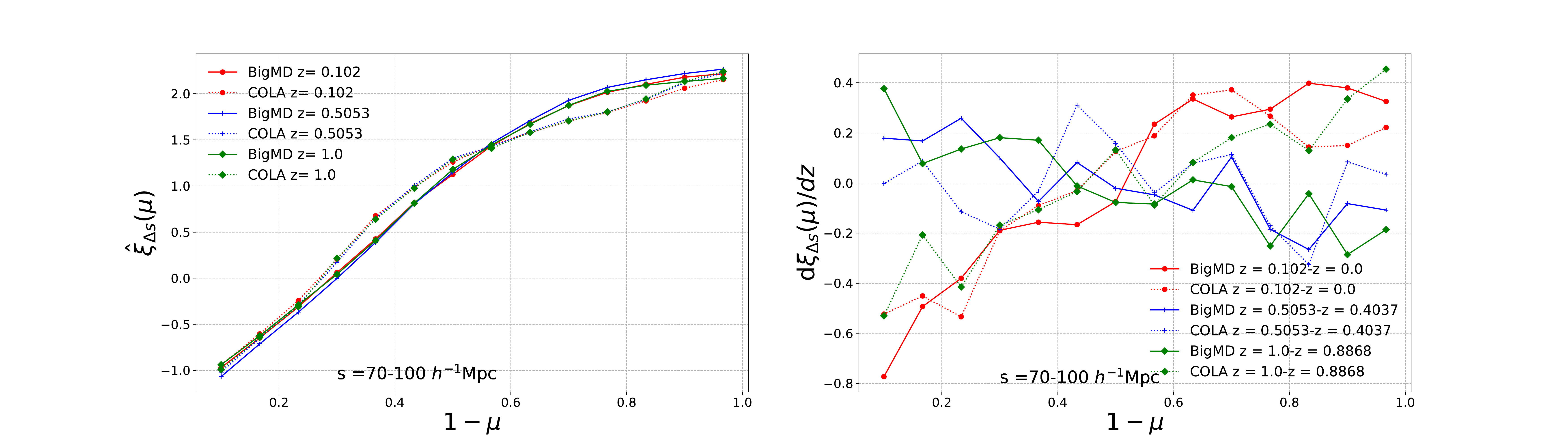}

  \caption{To test the performance of the method on other scales, 
  these show the values of ${\hat \xi_{\Delta s}(\mu)}$ and 
  ${d \hat \xi_{\Delta s}(\mu)} /{ dz}$ in clustering regions 
  of 6-40, 40-70 and 70-100 $h^{-1}\rm Mpc$, respectively. 
  We find that using large clustering scales results in 
  larger slope and evolution of slope in $\xi_{\Delta s}(\mu)$. 
  The statistical fluctuation (scattering)  increases with increasing clustering scales. }
  \label{fig_diffsbin}
\end{figure*}

\subsection{2pCF and anisotropy}

We use the 2pCF statistics to measure the anisotropies in the LSS clustering.
The Landy-Szalay estimator \citep{Landy} is adopted to calculate the 2pCF,
\begin{equation}
\xi(s,\mu) = \dfrac{DD-2DR+RR}{RR},
\end{equation}
where $DD$ is the number of galaxy-galaxy pairs, $DR$ is the number of galaxy-random pairs, 
and $RR$ is the number of random-random pairs.
All those numbers are measured with a dependence of $(s,\ \mu)$, 
where $s$ is the distance between the galaxy pair and $\mu = \cos(\theta)$, 
with $\theta$ being the angle between the line joining the pair of galaxies and the LOS direction. 
We use publicly available code \textsc{CUTE} \citep{CUTE} for computation of the 2pCF.  

In order to probe anisotropy, as was done in \cite{LI15,LI16}, 
we integrate the 2pCF over $s$ and only focus on the dependence on $\mu$,
\begin{equation}
\xi_s (\mu) =\int_{s_{\rm min}}^{s_{\rm max}} \xi(s,\mu) ds.
\end{equation}
The integration is, by default, conducted using $s_{\rm min} = 6 h^{-1}\rm Mpc$ and $s_{\rm max} = 40 h^{-1}\rm Mpc$
due to reasons explained in \cite{LI15,LI16}.
To remove the uncertainty from the clustering strength and {\it galaxy bias}, 
we further normalize the amplitude of $\xi_{\Delta s}(\mu)$ to only study its shape, i.e.
\begin{equation}\label{eq:norm}
\hat\xi_{\Delta s}(\mu) \equiv \frac{\xi_{\Delta s}(\mu)}{\int_{0}^{\mu_{\rm max}}\xi_{\Delta s}(\mu)\ d\mu}.
\end{equation}
A cut $\mu<\mu_{\rm max}$ is imposed to reduce FOG effect \citep{Jackson} 
(and also the fiber collision effect when studying the observational data).

\subsection{The Redshift Evolution} 

In addition to the AP effect, 
another source of apparent anisotropy in galaxy distribution we observed is the RSD effect due to the 
peculiar velocity of galaxies,
resulting in significant anisotropy even if the adopted cosmology is correct. 
\cite{LI14} found that, the anisotropy generated by the RSD effect is, although large, 
maintaining a nearly constant magnitude within a large range of redshift, 
while the anisotropies generated by AP varies with the redshift much more significantly.
So they proposed to measure the AP effect using the redshift dependence of the distortion.

Due to the growth of structure,
the galaxy peculiar velocities evolve with redshift and thus the RSDs are not exactly constant with time. 
The small redshift evolution of RSD would cause redshift-dependence in the LSS anisotropy,
which is the main source of the systematics of this method.

In this work, we measure $\delta \hat\xi_{\Delta s}(\mu,z_i,z_j)$ 
from the BigMD and COLA simulations to quantify the systematics from the RSD effect. 
The redshift evolution of the clustering anisotropy can be described as
\begin{equation} \label{eq:deltahatxi}
\delta \hat\xi_{\Delta s}(\mu,z_i,z_j)\ \equiv\ \hat\xi_{\Delta s}(\mu,z_i) - \hat\xi_{\Delta s}(\mu,z_j),
\end{equation}
where $z_i$ and $z_j$ are two different redshifts.
The above quantity then represents the systematics when the measurement is done in the correct cosmology (the simulation cosmology).
We further use 
\begin{equation}
 \frac{d \hat \xi_{\Delta s}(\mu)} { dz} \equiv \frac{\xi_{\Delta s}(\mu, z+\Delta z)-\xi_{\Delta s}(\mu,z-\Delta z)}{2\Delta z}
\end{equation}
to quantify the magnitude of the systematics contributed at a given redshift.


	
\begin{figure*}
	\centering
	\includegraphics[width=0.8\textwidth]{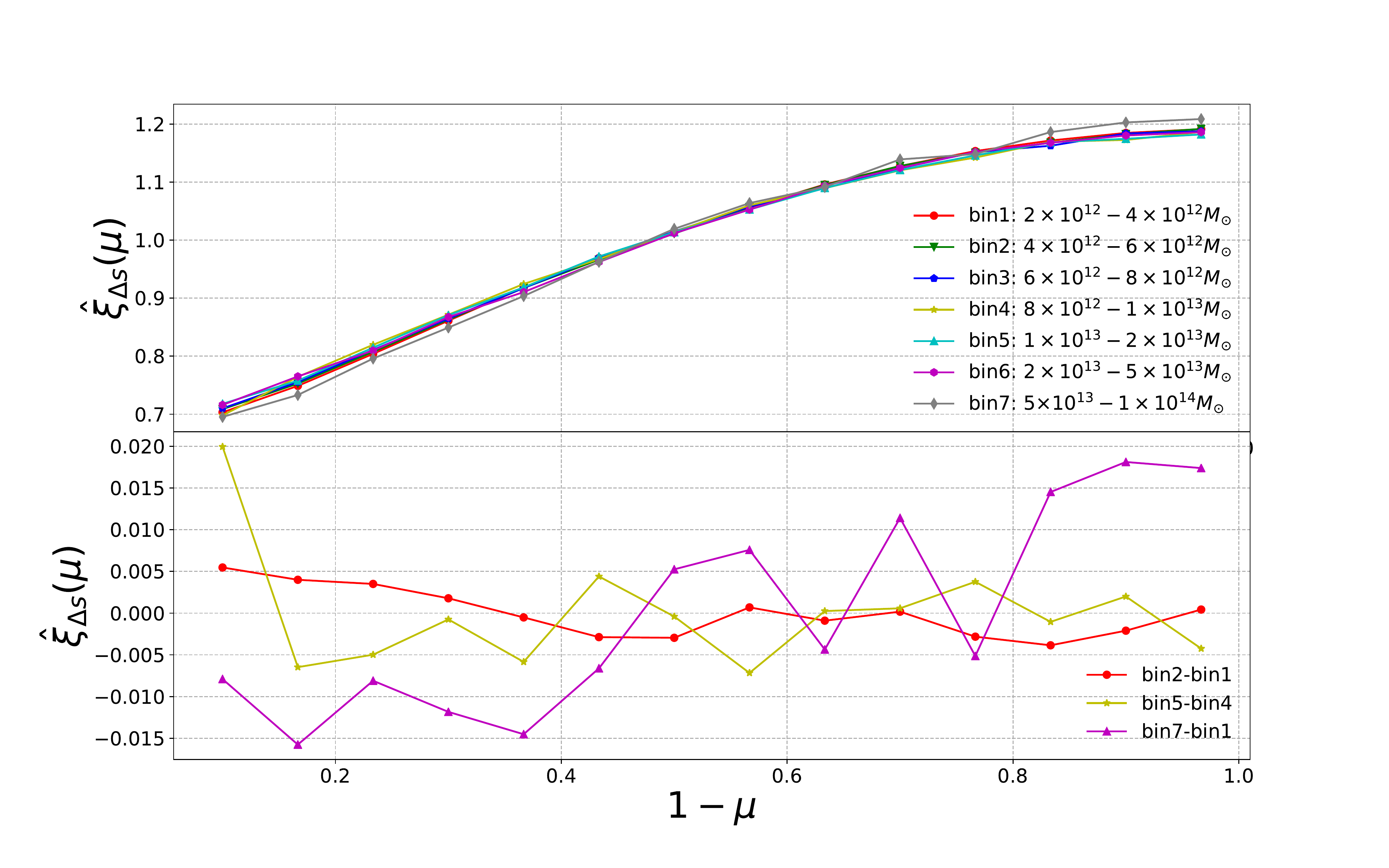}
	\caption{To investigate the effect of halo bias on 
	$\hat \xi_{\Delta s}$, 
	the upper panel presents the $\hat \xi_{\Delta s}(\mu,z=0)$,
	measured in 7 mass bins ranging from 
	$2\times 10^{12}M_{\odot}$ to $10^{14}M_{\odot}$. 
	We find that, the difference among them are inapparent, 
	mostly below $0.5$\%. 
	For the most extreme case, the results of using 
	$5\times10^{13}M_{\odot}<M<10^{14}M_{\odot}$ 
	is different form the $2\times10^{12}<M<4\times 10^{12}M_{\odot}$ 
	results on levels of $\approx$1.5\%.
	}
	\label{fig_massbin}
\end{figure*}

\section{Results} \label{sec:results}

In this section, we present the redshift evolution of $\hat \xi_{\Delta s}(\mu)$ measured from the BigMD and COLA simulations 
at redshift region of 0-1.445.
A comparison was made between the results measured from the two sets of simulations.

\subsection{Results of $\xi(s,\mu)$}


Figure \ref{contour} shows the 2D contour of the 2pCF $\xi(s,\mu)$, 
from the BigMD simulation and COLA simulation and at reshifts 0 and 1, respectively.
Due to the RSD effect, the contour lines are not horizontal.
The tilts at the left and right sides are caused by the FOG (finger-of-god) \citep{Jackson} 
and Kaiser effects \citep{kaiser1987clustering}, respectively.

The similarity between the results from the BigMD and COLA simulations 
implies that COLA as an approximate algorithm 
can successfully reproduces the clustering pattern
at $\gtrsim5 h^{-1}\rm Mpc$. 
The difference between the two sets of results are below a few percents.
Especially, the similarity in the region of $1-\mu\lesssim0.2$
suggests that COLA has the ability of correctly simulating the FOG effect.

 


\subsection{$\hat \xi(\mu)$ in the correct cosmology}


In Figure \ref{2pcf}, 
we compare the $\hat\xi_{\Delta s}(\mu)$s measured from the COLA and BigMD samples,
constructed using the underlying true cosmology (i.e., the simulation cosmology). 
We compute them by integrating $\xi(s,\mu)$ within 
the range $6h^{-1}{\rm Mpc}\le s \le 40h^{-1}{\rm Mpc}$, 
and normalizing the amplitude. 
The value of $\hat\xi_{\Delta s}$s are measured in 40 bins distributed in $0 \le \mu \le 0.94$.
Our results are as follows
\begin{itemize}
 \item The upper-left panels shows $\hat \xi_{\Delta s}$ measured at $z=0,\ 0.102,\ 0.5053,\ 1.0,\ 1.445$,
which clearly shows the anisotropy produced by the RSD effect, resulting in 30\% tilt in all curves.
Due to the non-zero redshift evolution of RSD, all curves are more tilted at higher redshift.
We find  that the results from COLA and BigMD only have a $\lesssim 1\%$ difference,
except the $1-\mu<0.2$ region at $z=0.102$.
\item To better quantify the redshift evolution of the clustering anisotropy, 
in the middle panels we plot the evolution against $z=0$,
showing the non-zero values of $\hat\xi_{\Delta s}(\mu,z) - \hat\xi_{\Delta s}(\mu,z = 0.0)$ caused by the redshift evolution of the RSD effect.
In the region of small/large $1-\mu$,
i.e. the region parallel/perpendicular to the LOS, the evolution becomes as large as 5/2.5\% at $z=0.5$, and increases to 10/5\% at $z=1,\ 1.445$.
COLA can well reproduces the BigMD results at levels of $\lesssim0.5$\%,
except the case of $z=1$, where we find a 2-3\% discrepancy.
\item 
In the lower panels we plot ${d \hat \xi_{\Delta s}(\mu)} /{ dz}$
at the three redshifts of $z=0.051,\ 0.4545,\ 0.9434$ to measure the redshift evolution from a different angle.
The estimation from COLA has $\lesssim5\%$ error at $z<0.5$,
and has a relatively large error of $15\%$ at $z\sim1$.
\end{itemize}





\subsection{$\hat \xi(\mu)$ in the wrong cosmologies}




In Figure \ref{wrongcos}, we plot the measured $\Delta \hat \xi_{\Delta s}$ 
and $d \hat \xi_{\Delta s} / dz$  
when assuming the "true" cosmology as well as a wrong cosmology $\Omega_m=0.25$, $w=-1.0$.
In the wrong cosmology, the amount of difference between COLA and BigMD results 
is similar to that which was measured in the correct cosmology, 
while this amount of difference is smaller than the shift of $\hat \xi_{\Delta s}$ and $d \hat \xi_{\Delta s} / dz$ induced 
by the wrong cosmology.
Clearly, COLA is accurate enough for the purpose of systematics estimation
in distinguishing this cosmology with the underlying true cosmology.


\subsection{Different clustering scales}


In previous studies of \cite{LI15,LI16,LI18,LI19} 
the authors only focus on the clustering scales of 
$6h^{-1}{\rm Mpc}\le s \le 40h^{-1}{\rm Mpc}$.
It is worthy testing the performance of the method on other scales.

In Figure \ref{fig_diffsbin} we plot the values of 
${\hat \xi_{\Delta s}(\mu)}$ and ${d \hat \xi_{\Delta s}(\mu)} /{ dz}$
in clustering regions of 6-40, 40-70 and 70-100 $h^{-1}\rm Mpc$, respectively.
We find that using large clustering scales results in 
larger slopes in $\xi_{\Delta s}(\mu)$. 

The $\gtrsim40h^{-1}\rm Mpc$ clustering scales are not used in the cosmological analysis of \cite{LI15,LI16,LI18,LI19} due to two reasons. 
Firstly, the statistical error increases as the clustering scale increases. 
Including their contribution into the integration $\int_{s_{\rm min}}^{s_{\rm max}} \xi(s,\mu) ds$
would not enhance the power of the statistics,
but rather decrease the S/N ratio and hence weaken the power of constraints.
Secondly, on large clustering scales, one can directly compare the theoretical expectation and the measured results of the 2pCFs,
to achieve a more complete exploration of the information (not just the AP signal) encoded in the data. 
The tomographic AP method is designed to explore the non-linear clustering regime
which is difficult for the traditional approaches. 




\subsection{Bias effect}

The effect of halo bias on $\hat \xi_{\Delta s}$ has been briefly investigated
in \cite{LI16}. Here we revisit this issue by using finer mass bins covering a larger range.

Figure \ref{fig_massbin} shows the $\hat \xi_{\Delta s}(\mu)$
at $z=0$, measured in 7 mass bins ranging from 
$2\times 10^{12}M_{\odot}$ to $10^{14}M_{\odot}$.
We find that, the difference among them are inapparent, mostly below $0.5$\%.
For the most extreme case, 
the results of using $5\times 10^{13}M_{\odot}<M<10^{14}M_{\odot}$
is different form the $2\times 10^{12}M_{\odot}<M<4\times 10^{12}M_{\odot}$ results
on levels of $\approx$1.5\%.


\section{Conclusion \label{sec:conclusion}}

The tomographic AP test is a novel statistical method 
entering nonlinear clustering scales of $6-40\ h^{-1}\rm Mpc$.
Tight cosmological constraints have been achieved 
by applying the method to the SDSS data \citep{LI16,LI18,LI19,Zhang2019}. 
The method measures the redshift evolution of anisotropy, 
usually quantified by $\hat \xi_{\Delta s}(\mu)$, 
to mitigate the effects from the RSDs while 
still being sensitive to the AP signal.
In this work, we studied the systematics of this methods in details, 
as an early step of its preparation of its application to future LSS surveys.

Since the next generation galaxy surveys will reach 
much higher redshift then the current ones, 
we study the redshift evolution of 
$\hat \xi_{\Delta s}(\mu)$ to the $z=1$ region.  
We find persistent redshift evolution in the whole region,
which manifests itself as a persistent, 
non-zero $d\hat\xi_{\Delta s}/dz$ existing 
within the whole redshift region.
Using $\hat\xi_{\Delta s}(\mu,z) - \hat\xi_{\Delta s}(\mu,z = 0.0)$ 
to quantify the difference between the high redshift and $z=0$ results,
we find an evolution of $\approx$5\% at $z=0.5$. 
The evolution reaches $\approx$10\% at $z=1$.

We also study the possibility of estimating the systematics 
via the cheap mocks generated by COLA.
In most cases we studied, COLA can well reproduces the N-body results.
In terms of estimating 
$\hat\xi_{\Delta s}(\mu,z) - \hat\xi_{\Delta s}(\mu,z = 0.0)$, 
COLA can reach an accuracy of $0.5\%$ at $z=0,0.5$,
while $2-3\%$ discrepancy is found at $z=1,1.445$.
Finally, we studied the dependence of the performance 
on the clustering scales and galaxy bias.

Our investigation suggests that the application 
of the method should be easy for $z\lesssim0.5$,
where the systematics are relatively small, 
and also can be accurately estimated using fast simulations.
In the next few years, surveys such as DESI  
and Taipan \footnote{https://www.taipan-survey.org/}  will probe the nearby universe 
with unprecedented precision.
Their datasets would be ideal for the application of our method.

Applying the method to region of $z\approx1$ would be challenging.
The systematics reaches $\approx10\%$,
while the inaccuracy of COLA also reaches $2-3\%$. 
To be well prepared for the cosmological analysis 
on the DESI, EULCID, WFIRST and PFS surveys,
this problem has to be resolved.
To achieve a fast and accurate systematics estimation,
one can improve the performance of the simulation by 
increasing the timestep, setting high initial redshift, 
trying alternative algorithms,  
or can make a correction to the COLA results using N-body results as baseline reference to decrease its inaccuracy.

This work shows that COLA is a possible tool for estimating systematics in multi-cosmologies. 
A fast and accurate enough systematics estimation method is important if one wants 
to conduct systematics correction in multiple cosmologies. 
In all previous works \citep{LI15,LI16,LI18,LI19}, 
the study of systematics was done using a simulation 
running under only one set of cosmological parameters.
The cosmological dependence of the systematics can be efficiently studied using the COLA algorithm.

A possible way to further reduce systematics is to design better statistics that is more insensitive to the redshift distortions. 
For example, we can use the marked correlation function \citep{Beisbart:2000ja,Gottloeber:2002vm,
Sheth:2004vb,Sheth:2005aj,White:2016yhs,Satpathy:2019nvo}, 
which assigns different weights to regions with different density, 
to suppress the strong RSD coming from the most clustered regions. 
These issues have not yet been discussed in details in this paper and are worth 
further investigating in future research.

In general, the tomographic method requires precise redshift measurement from spectroscopic surveys.
The method uses clustering region at $\approx5 h^{-1}{\rm Mpc}$,
requiring a redshift measurement with accuracy of $\delta z\approx0.001(1+z)$, 
which is difficult for most photometric surveys (e.g. DES and LSST).


The tomographic AP method is among the best methods in deriving cosmological constraints
using the $\lesssim 40h^{-1}\rm Mpc$ clustering region.
We will continually work on this method, so that we can safely use it to 
derive tight, robust cosmological constraints from the next generation LSS surveys.







\

\

\section*{acknowledgments}


We thank Kwan-Chuen Chan for suggesting us using the COLA algorithm.
XDL thanks Yi Zheng for helpful discussions on the systematics of the tomographic AP method. 
QM and YG thank Yuanzhu Huang and Yizhao Yang for many kind helps.
XDL acknowledges the supported from NSFC grant (No. 11803094).
ZL was supported by the Project for New faculty of Shanghai JiaoTong University
(AF0720053),
the National Science Foundation of China (No.  11533006, 11433001)
and the National Basic Research Program of China (973 Program 2015CB857000).
CGS acknowledges financial support from the National Research Foundation of Korea
(NRF; \#2017R1D1A1B03034900, \#2017R1A2B2004644 and \#2017R1A4A1015178).
This work was supported by World Premier International Research Center Initiative (WPI), MEXT, Japan.

We thank the CosmoSim database used in this paper, which is a service by the Leibniz-Institute for Astrophysics Potsdam (AIP).
We acknowledge the MultiDark database which was developed in cooperation with the Spanish MultiDark Consolider Project CSD2009-00064.

\appendix

\section{Comparison between the simulation and observational samples}\label{sec:RBtest}

As we are discussing ways of controlling the systematics of the tomographic AP method, 
the mock samples used in the analysis must be similar to those obtained in the observations.

\begin{figure}[H]
	\centering
	\includegraphics[width=1\textwidth]{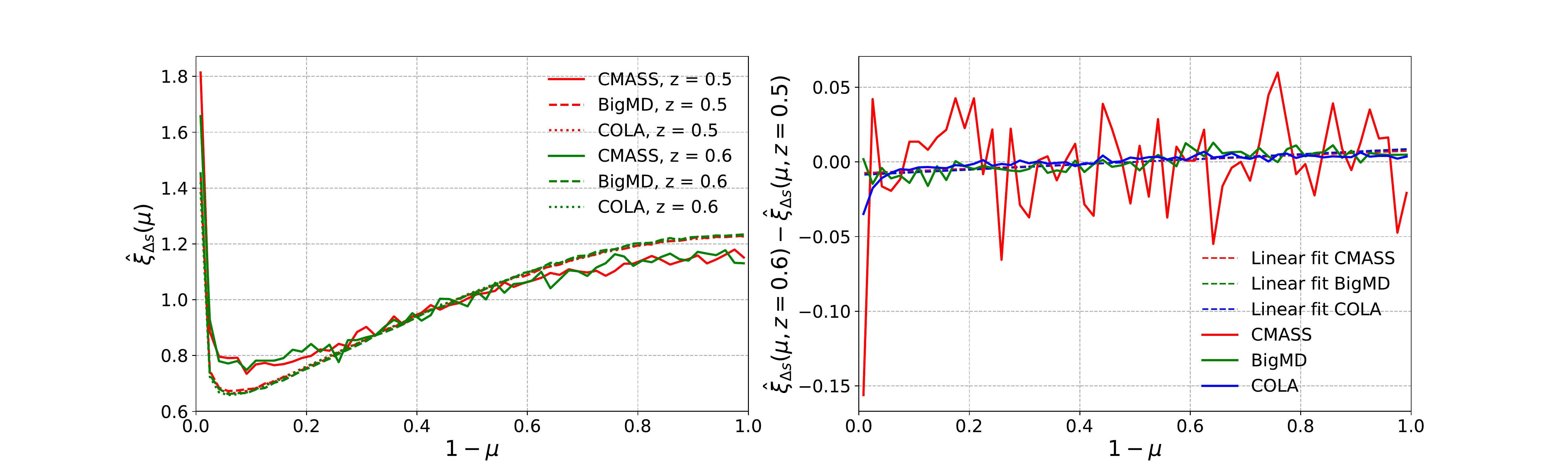}
	\caption{A comparison between the $\hat\xi_{\Delta s}(\mu)$ and its redshift evolution, 
	measured from the BigMD, COLA samples and the BOSS DR12 CMASS galaxies
	at redshifts of $z=0.48$ and $0.62$, respectively.
	The simulation samples can well reproduce the observatinoal measurements.}
	\label{fig:cmass_bigmd}
\end{figure}

Figure \ref{fig:cmass_bigmd} illustrates a comparison between the 2pCFs measured from 
1) two sets of BigMD \textsc{ROCKSTAR} halo samples, at the $z=0.48$ and $0.62$ snapshots;
2) two sets of COLA \textsc{ROCKSTAR} halo samples, at the same redshifts;
3) two sets of subsamples selected from the BOSS DR12 CMASS galaxies, by imposing the redshift range of $0.430<z<0.511$ and $0.572<z_6<0.693$;
these two subsamples have effective redshifts of $0.48$ and $0.62$, respectively.

In the left panel we plot the $\hat\xi_{\Delta s}(\mu)$s. 
The simulation samples achieve $\lesssim5\%$ level of accuracy in recovering the observational measurements.
The only exception is the leftmost point at $1-\mu=0$.
Here 1) the FOG effect is very strong; 2) the observational result may be problematic due to the fiber collision effect.
Anyway, this region is always abondoned when conducting cosmological analysis (\cite{LI16} imposed a cut $1-\mu>0.03$). 

Notice that the strong peak near $1-\mu=0$, produced by the FOG,
is not detected in Figure \ref{2pcf},\ref{wrongcos},\ref{fig_diffsbin},\ref{fig_massbin},
as we imposed a cut $1-\mu<0.03$ in these figures. 

The redshift evolution of $\hat\xi_{\Delta s}(\mu)$s is plotted in the right panel.
The observational results have much larger statistical error.
A linear regression shows that results from the BigMD, COLA and BOSS samples are in good consistency with each other.

In all plots, the BigMD and COLA results (the dashed and dotted curves) 
are so close to each other that, 
we can hardly distinguish them by eye. 

\bibliographystyle{aasjournal}

\end{document}